\documentclass[pre,twocolumn,showpacs,preprintnumbers,amsmath,amssymb,aps]{revtex4-1}

\usepackage{graphicx}
\usepackage{dcolumn}
\usepackage{bm}
\usepackage{amsmath} 
\usepackage{epstopdf}
\usepackage{epsfig}
\usepackage{subfigure}
\usepackage{fancyvrb}
\usepackage{natbib}
\setcitestyle{super}

\newcommand{\hamil}{\mathcal{H}}

\begin{document}


\title{Small-Network Approximations for Geometrically Frustrated Ising Systems}

\author{Bilin Zhuang}
 \altaffiliation[Current affiliation: ]{Division of Chemistry and Chemical Engineering, California Institute of Technology, Pasadena, CA 91125}
\author{Courtney Lannert}
 \email{clannert@wellesley.edu}
\affiliation{Department of Physics, Wellesley College, Wellesley, Massachusetts 02481, USA}
\date{July 14, 2011}

\begin{abstract}

The study of frustrated spin systems often requires time-consuming numerical simulations. As the simplest approach, the classical Ising model is often used to investigate the thermodynamic behavior of such systems. Exploiting the small correlation lengths in frustrated Ising systems, we develop a method for obtaining a first approximation to the energetic properties of frustrated two-dimensional Ising systems using small networks of less than 30 spins. These small networks allow much faster numerical simulations, and more importantly, analytical calculation of the properties from the partition function is possible. We choose Ising systems on the triangular lattice, the Kagome lattice, and the triangular Kagome lattice as prototype systems and find small systems that can serve as good approximations to these prototype systems. We also develop criteria for constructing small networks to approximate general two-dimensional frustrated Ising systems. This method of using small networks provides a novel and efficient way to obtain a first approximation to the properties of frustrated spin systems.   

\end{abstract}

\pacs{75.10.Hk, 75.10.-b, 75.10.Jm}
\maketitle

\section{Introduction}

The phenomenon of frustration in condensed-matter systems has been studied for over 50 years. In 1936, Pauling first noted that frustration is present in the structure of ice. The tetrahedral structure of ice allows multiple possible locations for the hydrogen atoms, giving rise to about $(3/2)^N$ ground-state configurations for a total of $N$ water molecules.~\citep{Pauling1935} Pauling's prediction was later confirmed by the experimental work by Giauque and co-workers.~\citep{Giauque1933, Giauque1936} Since then, frustration in condensed-matter systems has been an active area of research. Early theoretical investigations include the study of the antiferromagnetic triangular Ising lattice by Wannier, who showed that the system is disordered at all temperatures and the ground state entropy is $0.323Nk$, where N is the number of spins and $k$ is the Boltzmann constant.~\citep{Wannier1950} In recent years, new frustrated materials have been discovered and characterized following advancements in fabrication and measurement techniques. Most notably, the ground state spin configurations of rare earth pyrochlores $\text{Ho}_2\text{Ti}_2\text{O}_7$ and $\text{Dy}_2\text{Ti}_2\text{O}_7$ have been found to have a one-to-one correspondence to the structural configurations of ice, and thus they are aptly described as the ``spin ice".~\citep{Rosenkranz2000, Harris1997} Adding to their interest, Castelnovo et.\,al.\,recently proposed that magnetic monopoles emerge in the frustrated spin ice system,~\citep{Castelnovo2008} and this has been supported by a few recent experiments.~\citep{Morris2009, Bramwell2009, Fennell2009, Ladak2010} Apart from three-dimensional frustrated systems such as spin ice, two-dimensional frustrated systems with a range of different geometries can now be artificially constructed with nanometer-size magnetic islands \citep{Wang2006} or closely-packed colloidal spheres.\citep{Han2008} In addition, it has been proposed that frustrated spin systems can be constructed by trapping cold atoms~\citep{Duan2003} or polar molecules~\citep{Micheli2006} in an optical lattice. An important leap towards the experimental realization of such systems has been achieved by Simon et.\,al.\,by succesffully demonstrating antiferromagnetic spin chains in an optical lattice.~\citep{Simon2011}

Frustrated magnetic systems exhibit many interesting properties. Most notably, they have multiple degenerate ground states, which give rise to a non-zero entropy at absolute zero temperature, violating the third law of thermodynamics.~\citep{Ramirez2003} In addition, even in the regime when $kT$ is much less than the energy scale of the spin-spin interactions, there can still be significant fluctuations within the system.~\citep{Balents2010} Moreover, many frustrated systems possess rich phase diagrams as temperature or external magnetic field strength is varied. The phases displayed in these sytems can be magnetically ordered, partially ordered, or completely disordered.~\citep{DiepFrustrated} In addition to their theoretical interest, frustrated magnetic materials may have novel technological applications from microelectronics to drug delivery. For example, efficient and environmentally friendly magnetic refrigerators may be constructed with frustrated systems using the technique of adiabatic demagnetization~\citep{Zhitomirsky2003} and new technologies in advanced magnetic-recording devices may also be built with frustrated materials.~\citep{Wang2006} A better understanding of the behaviors of frustrated systems may even allow us to gain insights in fields beyond condensed matter physics. For instance, it has been suggested that the folding of a protein into a biologically funtionable structure is a result of the natural ability for the protein to resolve the frustrated couplings.~\citep{Bramwell2001}

As the simplest approach, the thermodynamic properties of geometrically frustrated systems can be studied with the Ising model. For systems in which classical spin fluctuations dominate, the Ising system can be a very accurate physical model. For other systems in which quantum fluctuations dominate, the Ising system becomes less accurate,~\citep{Balents2010} but we may still use it to obtain a first approximation. Currently, there are two primary methods for studying frustrated Ising systems:  exact analytical methods applied to infinite systems and Monte Carlo simulations applied on very large systems and extrapolated to the infinite-system limit. However, exact analytical solutions are not always possible, while Monte Carlo simulations can often be very time-consuming. Therefore, we are motivated to develop an efficient first approximation method for geometrically frustrated Ising systems.

One of the key features of geometrically frustrated systems is their small spin-spin correlation lengths. In the absence of long-range correlations, many properties of the system are determined by the local geometric network in the vicinity of each spin. By arranging a small number of Ising spins in an appropriate network, we find that the energetic properties of the extended two-dimensional Ising systems can be reproduced with surprising accuracy. Not surprisingly, the shorter the correlation length in the extended system, the better it can be approximated by a small network. There are two main advantages of using such small-network approximations. Firstly, since the small networks that we develop have less than 30 spins in general, it is much more efficient to do Monte Carlo simulations on the small networks. Secondly, for networks of less than about 25 spins, it is possible to calculate the thermodynamic properties of the system directly by utilizing the Boltzmann distribution with minimal computational power. This method of small-network approximation may provide us with a new and efficient first approximation to the properties of frustrated systems. When an appropriate small network is used, both exact analytical methods and Monte Carlo simulations are much simpler than for the associated extended Ising systems.

In this paper, we present an approximation technique for frustrated Ising systems; our goal is to find small Ising networks with less than 30 spins that accurately approximate the energetic properties of extended Ising systems in the thermodynamic (infinite-size) limit. We illustrate this technique on the two-dimensional triangular, Kagome, and triangular Kagome lattices. Since the energetic properties of these systems can be well represented by the specific heat vs.\,$kT$ profiles, we attempt to find small networks that accurately approximate the specific heat of the extended systems and develop general criteria for constructing good small-network approximations. The paper is organized as follows. In Section \ref{ModelDescription} and \ref{Numerical}, we describe our models and the numerical techniques that we use to study the models. In Section \ref{ResultDiscussion}, we compare the specific heat of particular small networks to that of the extended Ising systems on the triangular Kagome lattice, the Kagome lattice and the triangular lattice. In Section \ref{criteria}, we elucidate the general criteria for constructing a small network that well-approximates the specific heat of the extended two-dimensional lattices, and in Section \ref{PBC}, we discuss the advantages for using small-network approximations over carrying out simulations on a small piece of the extended lattice with periodic boundary conditions. Finally, in Section \ref{conclusion}, we offer some concluding remarks.

\section{Model Description}
\label{ModelDescription}

\subsection{The Extended Ising Models}
\begin{figure}[htbp]
\subfigure[]{\includegraphics[width=4cm]{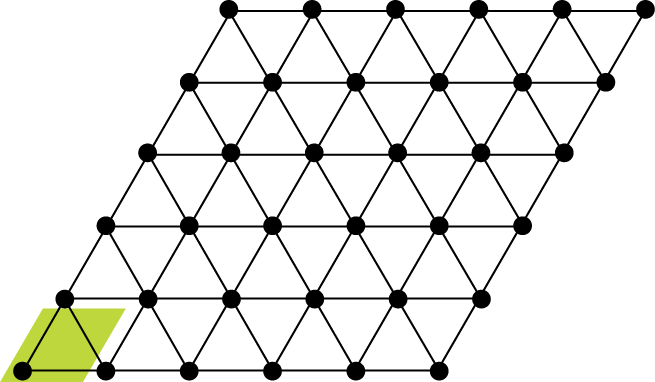}}
\subfigure[]{\includegraphics[width=3.3cm]{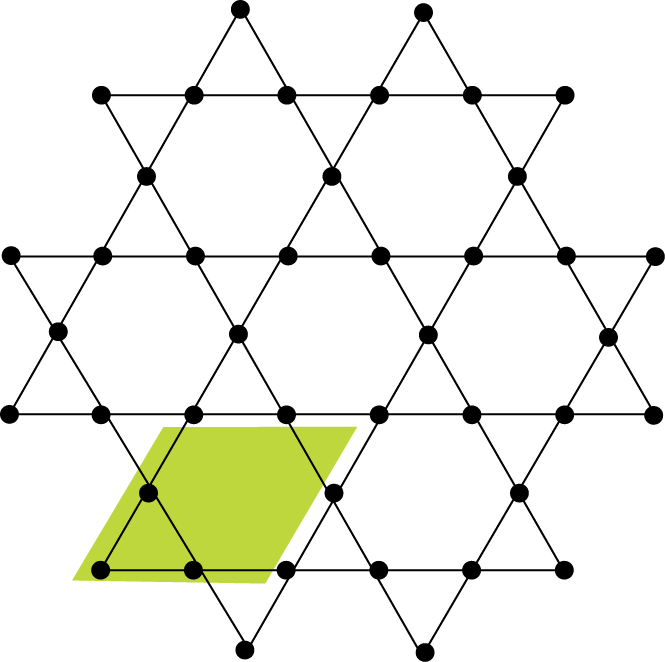}}
\subfigure[]{\includegraphics[width=4.5cm]{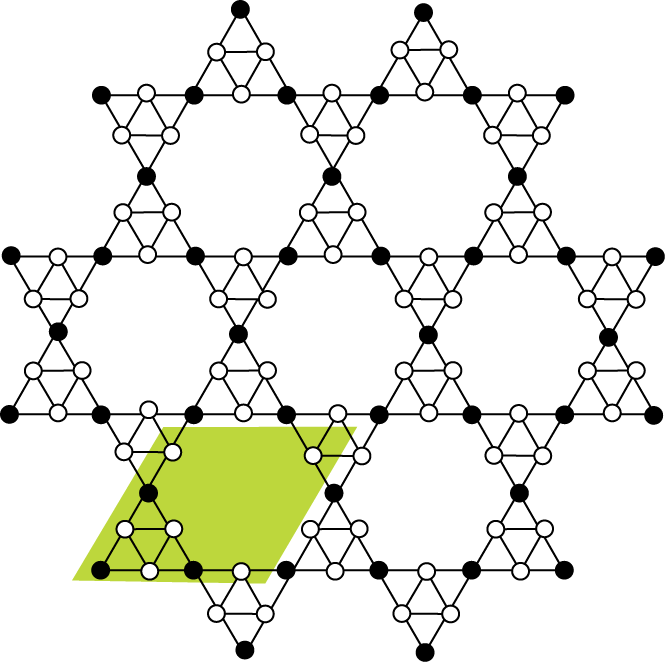}}
\caption{Schematic representations of the structures of ordinary two-dimensional Ising systems on (a) the triangular lattice, (b) the Kagome lattice and (c) the triangular Kagome lattice. The solid and open circles represent spin positions and the shaded areas represent a unit cell.}
\label{IsingSchematic}
\end{figure}

In this paper, we discuss the possibility of using small Ising systems to approximate the specific heat of geometrically frustrated two-dimensional Ising systems. We have chosen the Ising systems on the triangular lattice, the Kagome lattice and the triangular Kagome lattice as prototypes of extended two-dimensional systems. The network structures of these three lattices are schematically displayed in Fig.~\ref{IsingSchematic}, in which a solid or open circle represents a site for an Ising spin, and a single line represents the nearest-neighbor bond between the spins. We note that the triangular Kagome lattice is different from the other two lattices in that while all spins on the triangular lattice and the Kagome lattice are geometrically equivalent to one another, there are two kinds of sites with distinct geometries on the triangular Kagome lattice. We have distinguished the two kinds of spins using open and solid circles in Fig.~\ref{IsingSchematic}(c), and we shall refer to the spins represented by the open and solid circles as $a$-spins and $b$-spins respectively.

The Hamiltonian for an Ising system is given by~\citep{Ising1925}
\begin{equation}
\hamil=\sum_{\left\langle ij\right\rangle}J_{ij}\sigma_i\sigma_j
\label{Ising_ordinary}
\end{equation}
where the indices $i$ and $j$ label the lattice sites. $\sigma_i=\pm 1$ is the spin variable at site $i$, representing the \emph{up} and \emph{down} directions of the Ising spin, and $J_{ij}$ is the interaction between the spins on $i$ and $j$ sites. A positive value for $J_{ij}$ represents antiferromagnetic interactions while negative $J_{ij}$ represents ferromagnetic interactions. The summation runs over all nearest-neighbor pairs $\left\langle ij\right\rangle$.

In the case of the triangular lattice and the Kagome lattice, all spins are geometrically equivalent and all the nearest-neighbor interactions are of the same magnitude. Thus, $J_{ij}=J$ and we can rewrite the Hamiltonian as 
\begin{equation}
\hamil=J\sum_{\left\langle ij\right\rangle}\sigma_i\sigma_j
\label{Ising 0 h}
\end{equation}

In the case of the triangular Kagome lattice, there are two kinds of nearest-neighbor interactions: the $aa$-interaction and the $ab$-interaction. We denote the strength of the two kinds of interactions as $J_{aa}$ and $J_{ab}$ respectively, and the Hamiltonian of the system is given by
\begin{equation}
\hamil=J_{aa}\sum_{i,j\in a} \sigma_{a,i} \sigma_{a,j} + J_{ab}\sum_{i 
\in a, j \in b} \sigma_{a,i}\sigma_{b,j}
\label{Hamiltonian}
\end{equation}
where $\sigma_{x,i}$ denotes the spin with index $i$ on the $x$-lattice ($x=a,b$). The first summation runs over all nearest-neighbors $\left\langle ij\right\rangle$ within the $a$-sublattice and the second summation runs over all nearest-neighbors $\left\langle ij\right\rangle$ between the $a$- and the $b$-sublattices.

\subsection{The Small Ising Networks}

The small Ising systems that we have constructed to approximate the extended lattices in the previous section span a range of structures. Each consists of a group of less than 30 spins connected in a specific network. In many cases, the network of a particular small Ising system can be represented by a polyhedron, on which the vertices represent the spin positions and the edges represent the nearest-neighbor bonds. We will display the structures of the small systems as we discuss the small-network approximations in Sec.~\ref{ResultDiscussion}.

As in the extended Ising systems, each spin position on the small networks has a spin variable of $\sigma_i = \pm 1$, representing the \emph{up} and \emph{down} orientations of the spin. Since most of our small networks can be represented by three-dimensional structures such as polyhedra, the terms \emph{up} and \emph{down} may cause potential confusion due to the surface curvature of the structures. In our case, \emph{up} and \emph{down} represent the absolute up and down directions in spin space and not the directions perpendicular to the surface of the polyhedron at any particular spin position.

The Hamiltonian of a small network is given by
\begin{equation}
\hamil=\sum_{\left\langle ij\right\rangle} J_{ij} \sigma_i \sigma_j 
\label{HamiltonianSmall}
\end{equation}
where $J_{ij}$ is the interaction between spins on sites $i$ and $j$, and the summation runs over all the bonds in the network.

\section{Numerical Techniques}
\label{Numerical}

In this section, we introduce the numerical methods involved in our study. Section \ref{Section3A} gives the numerical details of the Monte Carlo simulation, which is used to calculate the properties of both the extended lattices and the small networks. In Section \ref{Section3B}, we present the method for calculating the specific heat exactly from the partition function; this method is computationally inexpensive for systems with less than about 25 spins. In Section \ref{Section3C}, we introduce a ``deviation index" to quantify how well each small network approximates the thermodynamics of the extended lattice system.

\subsection{Monte Carlo Simulation}
\label{Section3A}
Frustrated Ising systems exhibit little long-range order, and therefore a single-spin-flip algorithm is suitable to statistically sample the microstates of these systems. In view of this, we employ the Metropolis algorithm to study the Ising systems on both the extended two-dimensional lattices and the small networks.~\citep{Metropolis1953, LandauMonteCarlo} For ordinary extended two-dimensional Ising systems, we carry out the simulations using systems of $L \times L$ unit cells under periodic boundary conditions. We use $L=30$, $18$ and $10$ for the triangular lattice, the Kagome lattice, and the triangular Kagome lattice, respectively, so that there are at least $900$ spins in the simulation box for each system. Each simulation is started with a randomly generated spin configuration on the lattice, and 1000 Monte Carlo steps were performed to allow the system to equilibrate before any measurements were taken.  For a system of $N$ spins, one Monte Carlo step is equivalent to $N$ Metropolis loops. (So that in each step, each spin in the system is chosen once on average by the algorithm.) After the initial equilibration steps, one measurement of energy is recorded for each of the 9000 subsequent steps. 10 independent simulations are carried out for each extended lattice and the results are averaged.

For the small networks, we carry out 1000 equilibration steps followed by 9000 measurement steps as well. However, we average the result from 100 sets of Monte Carlo simulations so that our results are statistically significant. Because each of the 100 Monte Carlo simulations are independent of one another, it is possible to carry out the 100 sets of Monte Carlo simulations simultaneously with a single processor, which significantly reduces the computer time required for the simulation. 

In this paper, we are interested in calculating the energy and the specific heat of our systems. The energy can be calculated directly by finding the ensemble average of the Hamiltonian, which is just the averaged energy for all the microstates sampled:
\begin{equation}
E=\left\langle \hamil_\mu\right\rangle
\end{equation}
where the subscript $\mu$ denotes a sampled microstate.

On the other hand, the heat capacity of the system can be calculated directly from the energy by applying the fluctuation dissipation theorem on the system.~\citep{NewmanMonteCarlo} As the heat capacity is directly proportional to the total number of bonds in the system, we have to divide the heat capacity by the total number of bonds to allow a fair comparison between the various models. In our calculation, we define the specific heat, $c$, as the heat capacity per nearest-neighbor bond:
\begin{equation}
c=\frac{2}{Nz}\frac{1}{kT^2}\left( \left\langle \hamil_\mu^2 \right\rangle - \left\langle \hamil_\mu \right\rangle^2 \right)
\end{equation}
where $N$ is the number of spins in the system and $z$ is the coordination number for each spin. The factor of $2$ corrects for double counting.

Frustrated Ising systems that have a uniform interaction constant across the system undergo an ``excitation transition" as the temperature increases from zero. Below the temperature of the excitation transition, the system is in one of its ground states. At the excitation transition, the excited states become accessible to the system. The temperature of the excitation transition is largely dependent on the geometry of the Ising system, and we have found that the excitation transition of an extended frustrated Ising system can be modeled using our small networks. The behavior of the excitation transition can be best studied using the specific heat as a function of temperature, in which the excitation transition manifests as a round and broad peak. In addition, two systems with the same specific heat vs.\,temperature profile have the same energy profile as well. In view of this, we focus on using small networks to approximate the specific heat vs.\,$kT$ profiles of the extended systems.

\subsection{Analytical Calculation}
\label{Section3B}

As our small networks have a very limited number of spins, their specific heat may also be calculated using the canonical ensemble and considering all possible microstates for the system. In this formulation, the partition function is given by:
\begin{equation}
Z= \sum_{\text{all states}} e^{-\beta E_\mu}
\label{partitionZ}
\end{equation}
where the subscript $\mu$ designates each microstate and $\beta=1/kT$. The specific heat can be calculated from the partition function using the following expression:~\citep{HuangStatMech}
\begin{equation}
c=\frac{2}{Nz}k\beta^2\frac{\partial^2 \ln Z}{\partial \beta^2}
\label{cAnalytical}
\end{equation}

For very small systems of less than 25 spins, it is easy to numerically evaluate the partition function using Eq.~\eqref{partitionZ} and then calculate the specific heat using Eq.~\eqref{cAnalytical}. This provides us with an alternative way to calculate the specific heat of our small networks, and this exact method requires much shorter computation time than a Monte Carlo simulation. 

\subsection{The Deviation Index}
\label{Section3C}

In this work, we consider a number of small networks as approximations to the extended lattices under study. Some are better approximations to the corresponding extended lattice, while others are not as good. To evaluate how good a small-network approximation is, we define a deviation index, $D$, which is a generalization of the coefficient of determination, $R^2$, in statistics.~\citep{ProbR2} Usually, $R^2$ is used to determine the accuracy of a curve fit and has values between $0$ and $1$. The closer the value of $R^2$ to $1$, the better the approximation. However, in our case, the value of $D$ may go below zero when the two curves are very far apart.

Our simulations are performed for the same set of $kT/J$ values for each system. For the $n$th value of $kT/J$, we obtain the specific heat values $c_{\text{ext},n}$ and $c_{\text{sn},n}$ for the extended system and the small network, respectively. We then calculate the deviation index $D$ using the following equations:
\begin{eqnarray}
SS_\text{tot}&=&\sum_n (c_{\text{ext},n}-\bar{c}_{\text{ext}})^2 \\
SS_\text{err}&=&\sum_n (c_{\text{ext},n}-c_{\text{sn},n})^2 
\end{eqnarray}
and
\begin{equation}
D=1-\frac{SS_\text{err}}{SS_\text{tot}}
\label{R_sq}
\end{equation}
where $\bar{f}$ is the mean value of $f_n$.

To allow comparisons over a range of values of $kT/J$, we carry out our simulation for $150$ values of $kT/J$ equally spaced in logarithmic scale between $kT/J=0.1$ and $kT/J=100$. 

\section{Comparison between Small Networks and Extended Systems}
\label{ResultDiscussion}

Geometrically frustrated systems are notable for their short correlation lengths. As suggested by Table~\ref{tab_SCompare}, the spin-spin correlation length ranges from extremely small for the triangular Kagome lattice to power-law suppressed for the triangular lattice with the Kagome lattice intermediate with exponentially-suppressed correlations. It is this absence of long-range correlations that make it possible to approximate the collective behavior of an infinite system by a small number of spins connected in an appropriate network. In a suitable small network, the local interactions are preserved but the long-range correlations are neglected. Accordingly, we expect our small network approximation method to succeed for a particular extended system precisely to the degree that that system has short-ranged correlations.
\begin{table}[htbp]
\centering
\begin{tabular}{cc}
\hline
Lattice  & Spin-Spin Correlation\\
\hline
Triangular Lattice        & $C (r/r_0)^{-1/2}$ with $|C|\sim 1$\footnote{Reference\citep{Stephenson1964}}\\
Kagome Lattice            & $e^{-r/\xi}$ with $\xi=3.3r_0$\footnote{Reference\citep{Suto1981,Broholm1990}}\\
Triangular Kagome Lattice & 0 for $r/r_{bb} \geq 1$ \footnote{Reference~\citep{Loh2008}}\\
\hline
\end{tabular}
\caption{Spin-spin correlations of three antiferromagnetic lattices at zero temperature and zero field, where $r_0$ is the lattice constant for the triangular and Kagome lattcies and $r_{bb}$ is the distance between $b$ sites in the triangular Kagome lattice.} 
\label{tab_SCompare}
\end{table}

\subsection{The Triangular Kagome Lattice}

We begin by developing a small-network approximation for the frustrated triangular Kagome lattice (TKL) which has the shortest-range correlations of the three systems we consider. Unlike the triangular and the Kagome lattices, the TKL does not require all bonds to have $J>0$ to be frustrated. As long as $J_{aa}>\left|J_{ab}\right|$, the system is frustrated, remaining disordered down to zero temperature,~\citep{Loh2008} regardless of the sign of $J_{ab}$. In addition, earlier work by Loh et.\,al.\,has shown that the sign of $J_{ab}$ does not affect the energy and the specific heat of the system, due to a one-to-one correspondence between the microstates for two systems with the same $J_{aa}$ but with $J_{ab}$ differing by a sign.~\citep{Loh2008} Due to the two kinds of bonds in the TKL, there are in general two excitation transitions for the system, appearing as two broad peaks in the specific heat vs.\,$kT$ plot. By seeking to reproduce the interactions and the connectivities of the lattice sites in the TKL, we arrived at the ``triangular drying-rack network'', which we now show is a very good approximation to the frustrated TKL.

\begin{figure}[htbp]
\centering
\subfigure[]{\includegraphics[width=3cm]{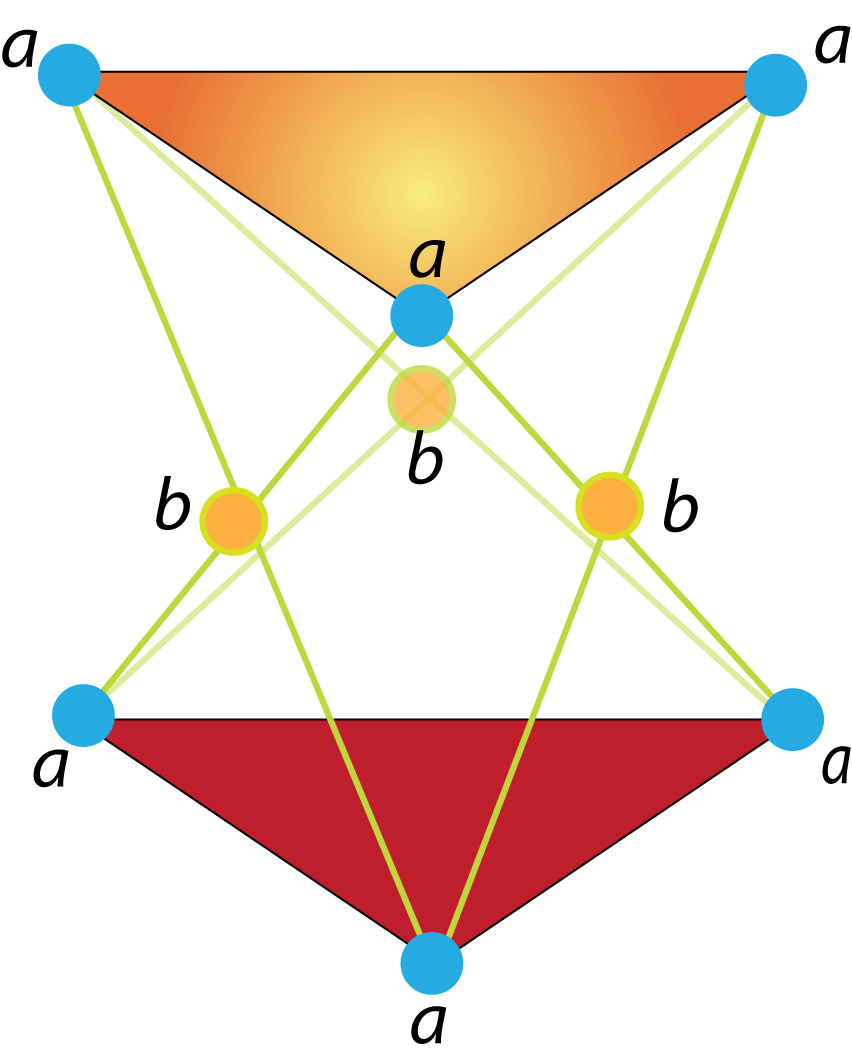}}\qquad
\subfigure[]{\includegraphics[width=4.5cm]{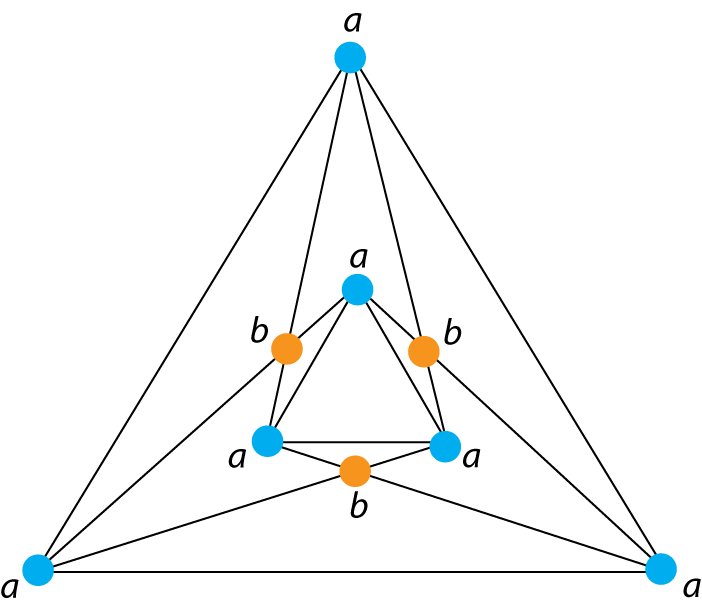}}
\caption{(a) Structure of the triangular drying-rack network; (b) Connectivity between spins in the triangular drying-rack network}
\label{TriDryingRack}
\end{figure}

A schematic diagram of our small-network approximation to the TKL is shown in Fig.~\ref{TriDryingRack}(a). This structure has 9 spins and 18 bonds in total. Two $a$-trimers are located on top and at the bottom of the ``drying-rack'', and three $b$-spins are located on the three sides, providing the connection between the top and the bottom trimers. Fig.~\ref{TriDryingRack}(b) displays the connectivity between the spins in the drying-rack, showing six $aa$-bonds and twelve $ab$-bonds in the structure. The ratio between the number of $aa$-bonds and $ab$-bonds in this structure is the same as that in the extended two-dimensional TKL.

\begin{figure}[htbp]
\centering
\subfigure[]{\includegraphics[width=9cm]{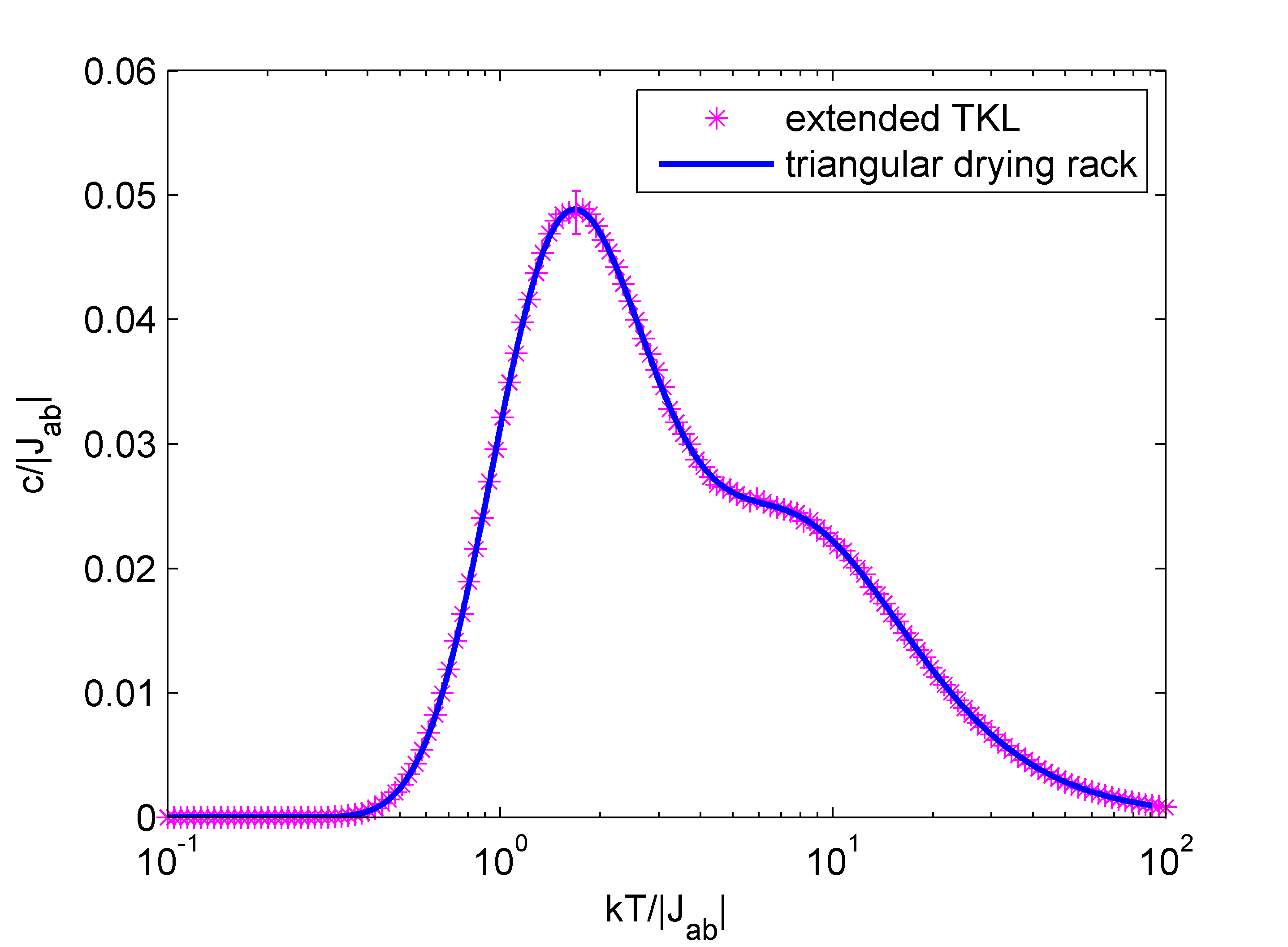}}\qquad
\subfigure[]{\includegraphics[width=9cm]{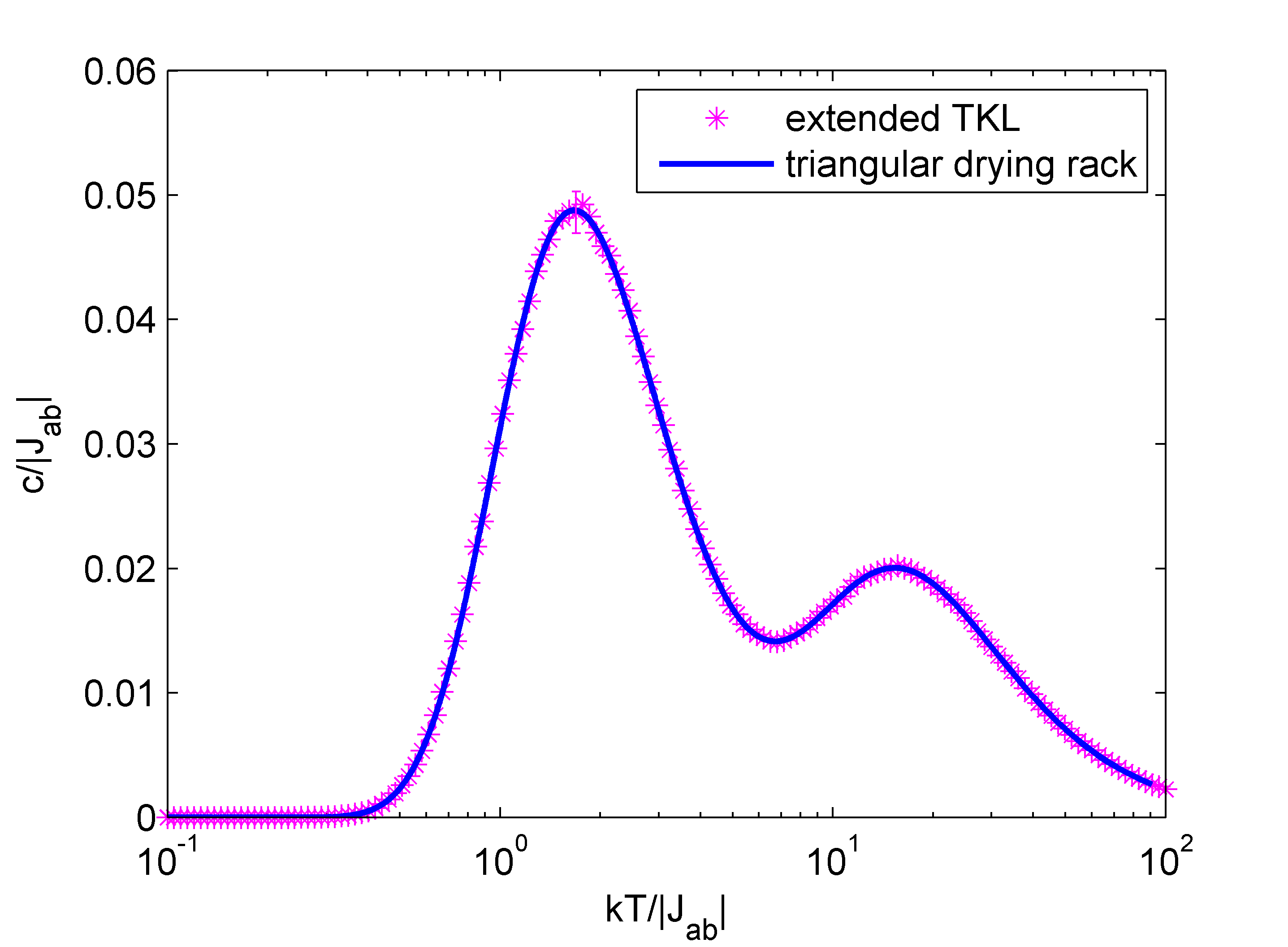}}
\caption{Specific heat vs.~$kT$ plots for the triangular drying rack network and the extended two-dimensional TKL with (a) $J_{aa}/\left|J_{ab}\right|=5$  and (b) $J_{aa}/\left|J_{ab}\right|=9$.}
\label{TKL_TDR_C}
\end{figure}

In each set of plots in Fig.~\ref{TKL_TDR_C}, we compare the specific heat vs.~$kT$ profiles for the triangular drying-rack network and the Ising model on the extended TKL with the same $J_{aa}$ and $J_{ab}$ values. Although the triangular drying-rack network has only 9 spins, its specific heat profile matches that of the extended TKL almost exactly. Since the specific heat profile of the full TKL can be modeled by such a limited number of spins in our triangular drying-rack network, it suggests that it is possible to study frustration by focusing on the local interactions among a small number of spins that are connected in an appropriate network.

\subsection{The Kagome Lattice}

\begin{figure}[htbp]
\centering
\subfigure[]{\includegraphics[width=3cm]{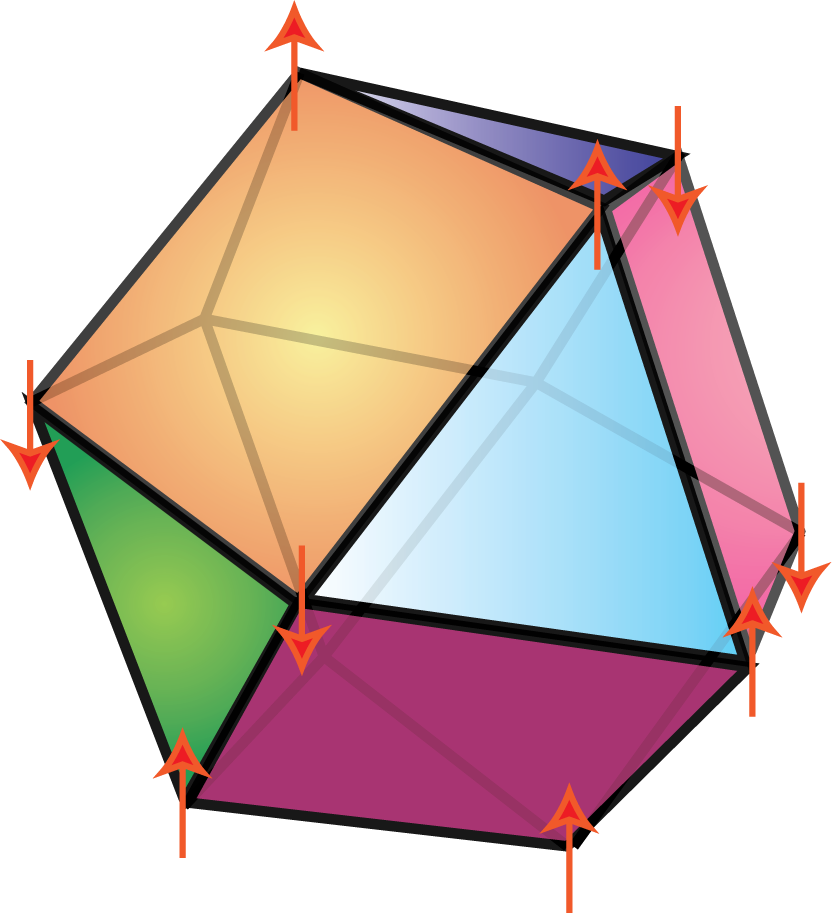}}\qquad
\subfigure[]{\includegraphics[width=3cm]{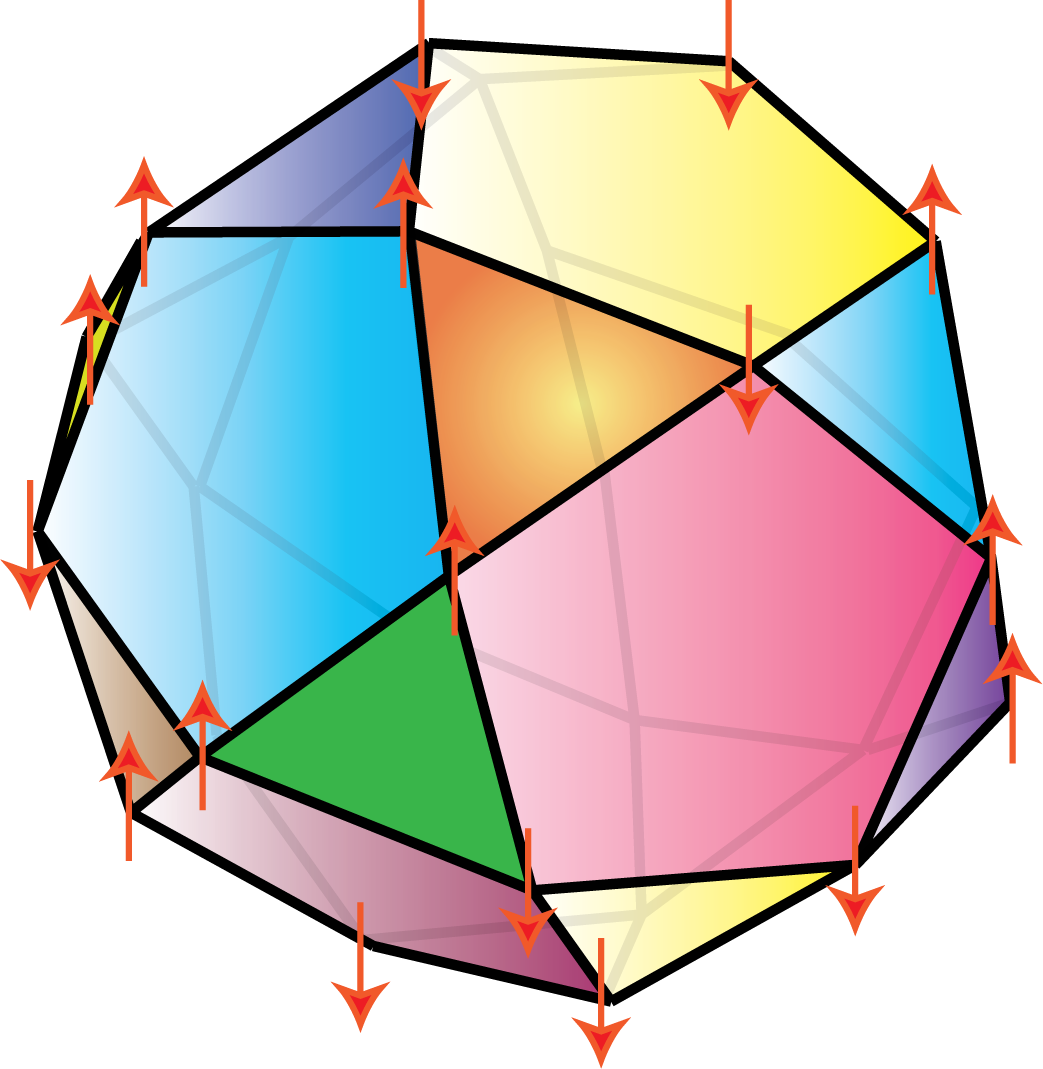}}
\caption{Structure of Ising systems on (a) a cuboctohedron, and (b) an icosidodecahedron.}
\label{cubo_icosi}
\end{figure}

In this section, we compare the specific heat profiles of particular small Ising networks with that of the frustrated Ising system on the Kagome lattice. In the Kagome Ising system, each spin has four nearest neighbors that are arranged in a bowtie network around the spin, and all spins on the lattice are geometrically equivalent to one another. In view of these geometrical features of the Kagome lattice, we consider Ising networks represented by two quasiregular polyhedra: the cuboctahedron and the icosidodecahedron. The structures of the cuboctahedron and the icosidodecahedron are shown in Fig.~\ref{cubo_icosi}(a) and (b) respectively. We treat each vertex on the polyhedra as a spin position and each edge as a bond between the neighboring spins. Here, the cuboctahedron has 12 spins and 24 bonds, while the icosidodecahedron has 30 spins and 30 bonds. Both small networks are similar to the Kagome lattice in the sense that each spin and its four neighbors form a bowtie network and all spins are geometrically equivalent to one another within the structure. However, while the Kagome lattice consists of triangular and hexagonal plaquettes, the cuboctahedron and the icosadodecahedron differ by containing square or pentagonal instead of hexagonal faces.

\begin{figure}[h]
\centerline{\includegraphics[width=9cm]{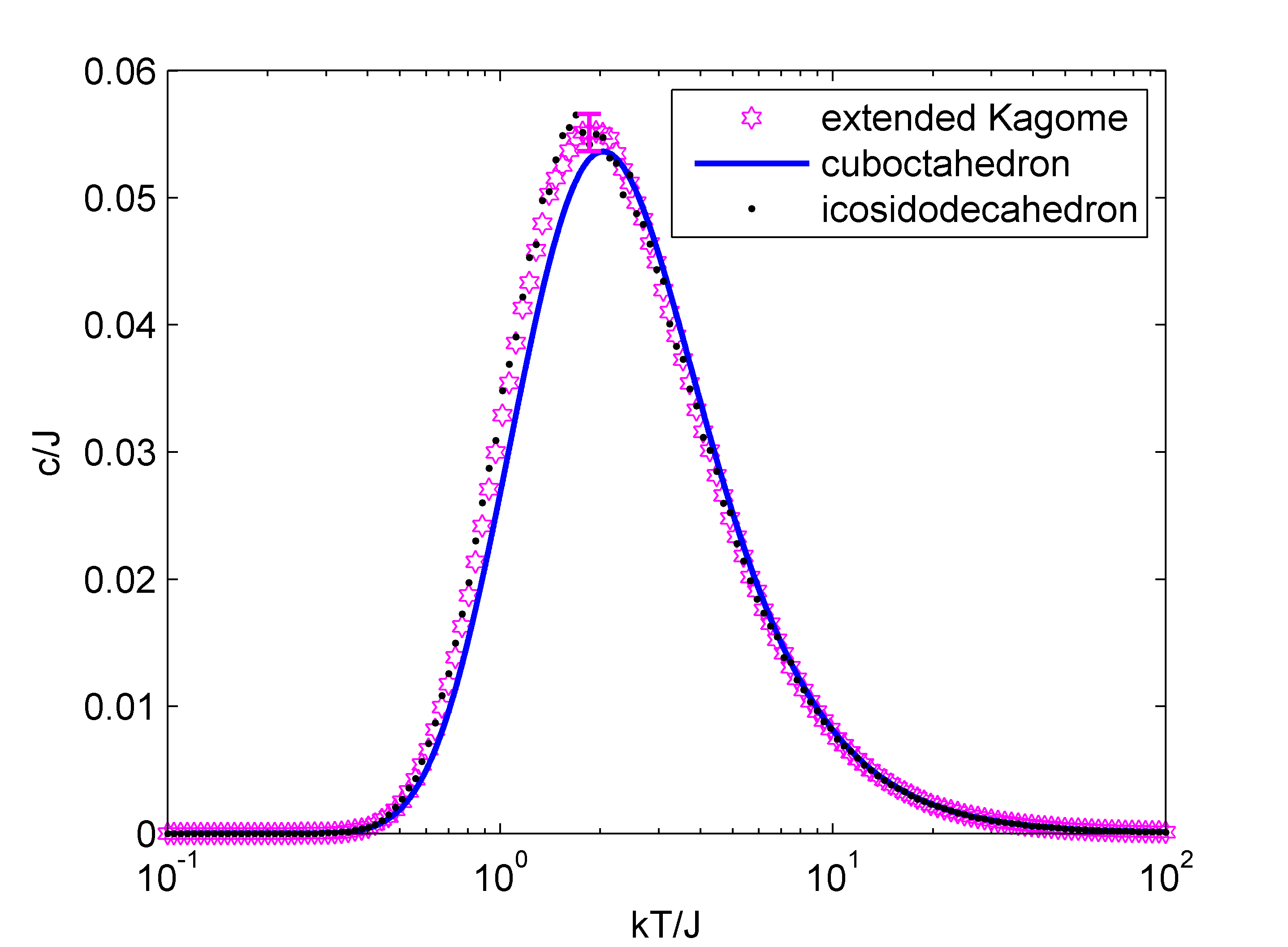}}
\caption{The specific heat vs.~$kT$ plot for frustrated Ising systems on the Kagome lattice, the cuboctahedron network and the icosidodecahedron network}
\label{C_cubo_icosi}
\end{figure}

The specific heat profiles for the Ising systems on the Kagome lattice, the cuboctahedron network and the icosidodecahderon network are shown in Fig.~\ref{C_cubo_icosi}. Comparisons show that the icosidodecahedron's network, which has only 30 spins, can serve as a very good approximation to the specific heat profile of the frustrated Kagome lattice at all $kT$. On the other hand, the cubotahedron network is a less accurate approximation, which produces a slightly lower peak than the extended Kagome lattice in the specific heat vs.\,$kT$ plot. However, the cubotahedron network has the advantage that it has only 12 spins, which makes it very easy to analytically evaluate the specific heat exactly using the partition function. In Fig.~\ref{C_cubo_icosi}, the result for the cuboctahedron network is obtained using the analytical method, while the result for the icosidodecahedron network is obtained from Monte Carlo simulation.

\begin{figure}[htbp]
\centering
\subfigure[]{\includegraphics[width=3cm]{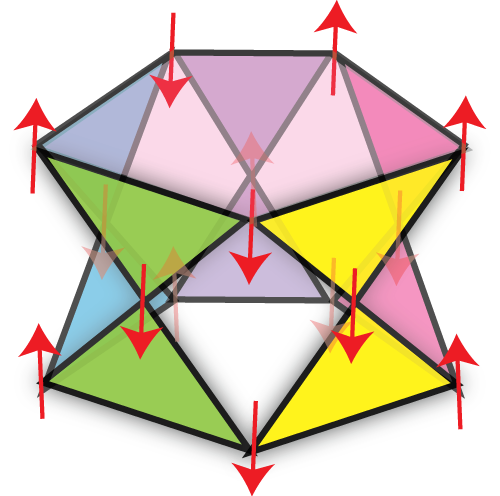}}\qquad
\subfigure[]{\includegraphics[width=3cm]{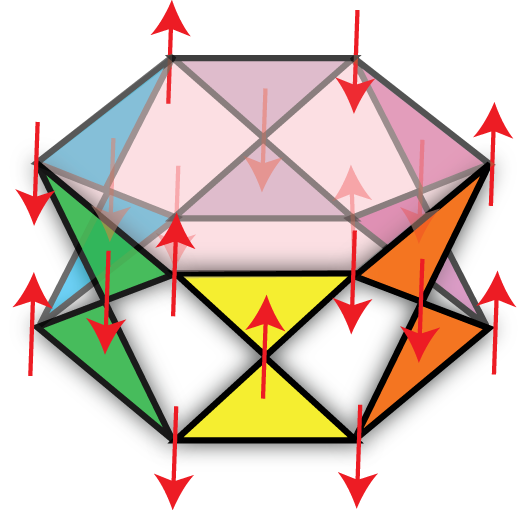}}
\caption{Structure of Ising systems on (a) the 5-bowtie network, and (b) the 6-bowtie network.}
\label{N-bowties}
\end{figure}

\begin{figure}[htbp]
\centerline{\includegraphics[width=9cm]{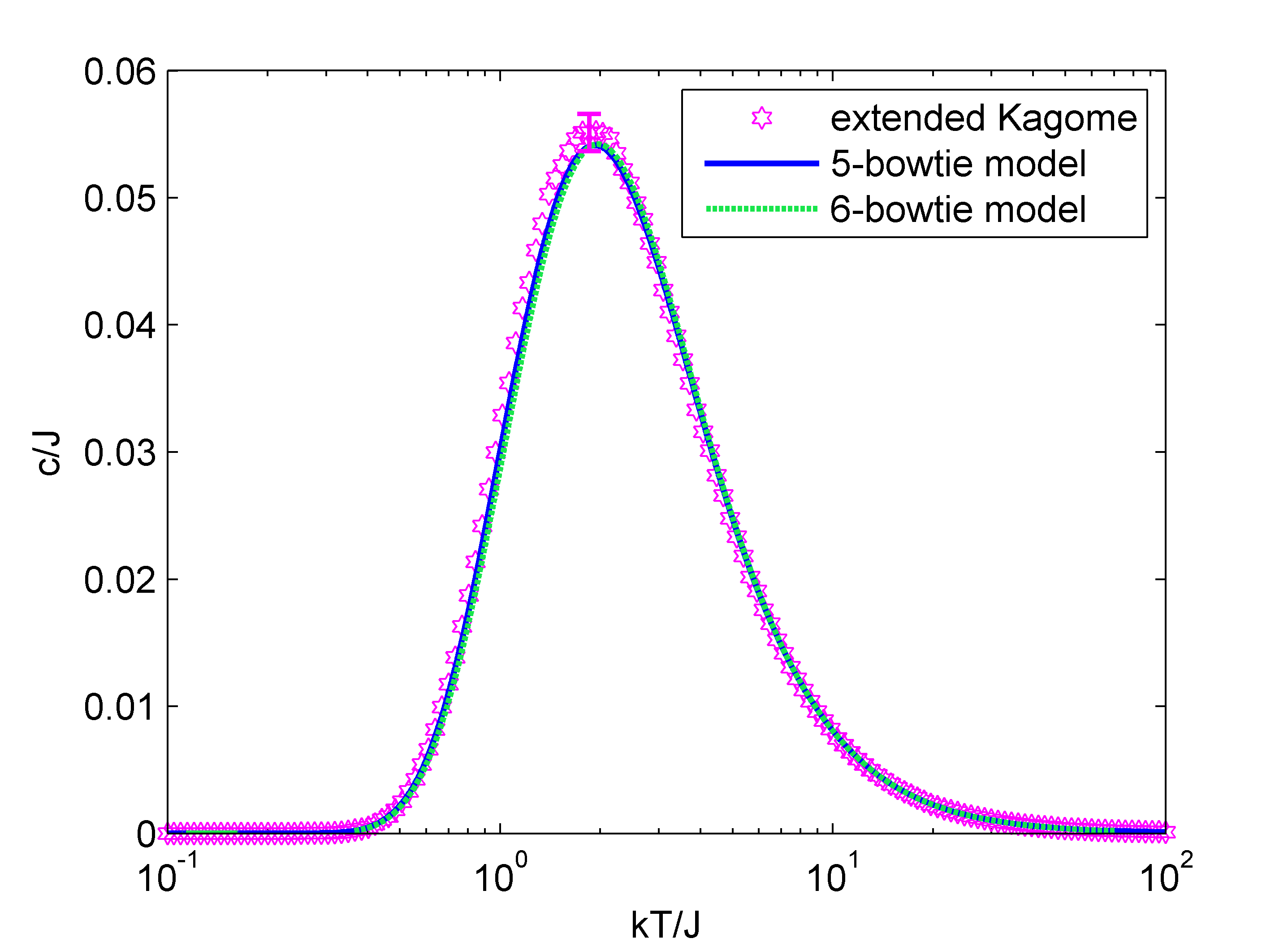}}
\caption{The specific heat vs.~$kT$ plot for frustrated Ising systems on the Kagome lattice, the 5-bowtie network and the 6-bowtie network}
\label{Heat_capa_bowtie}
\end{figure}

Both the cuboctahedron and the icosidodecahedron belong to the class of quasiregular polyhedra, as all the vertices are geometrically equivalent to one another on each polyhedron. The geometrical equivalence of all spins is similarly true for the Kagome lattice. However, we show next that it is possible to construct a small network that well-approximates the Kagome lattice without requiring geometrical equivalence between the sites. Consider two other networks, which we name the ``5-bowtie network'' and the ``6-bowtie network'', which are composed of five or six bowties wrapped around in a circle. The structures of these two networks are shown in Fig.~\ref{N-bowties}. In these two networks, not all spins are geometrically equivalent to one another. The specific heat vs.\,$kT$ curves for these bowtie networks in Fig.~\ref{Heat_capa_bowtie} show that they are also good approximations to the Kagome lattice.

\subsection{The Triangular Lattice}

\begin{figure}[htbp]
\centering
\subfigure[Tetrahedron]{\includegraphics[width=2.7cm]{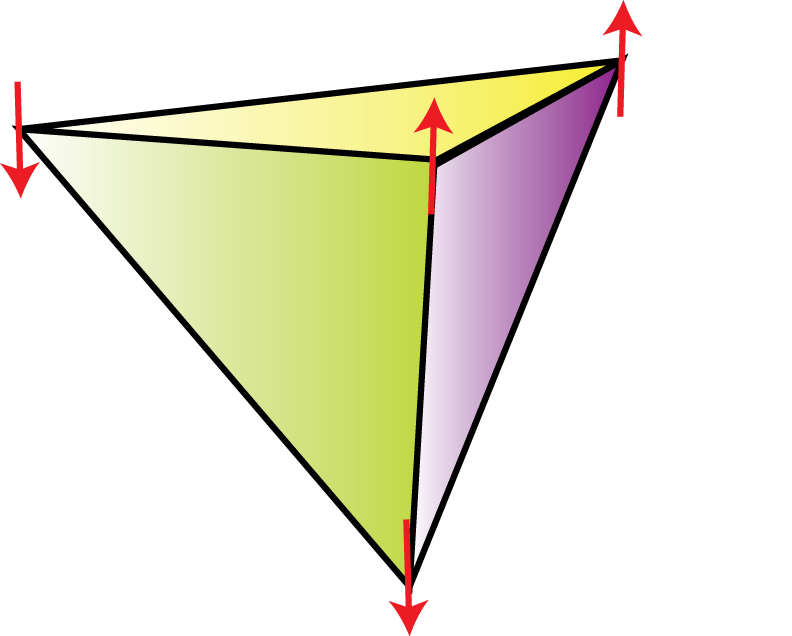}} 
\subfigure[Octahedron]{\includegraphics[width=2.5cm]{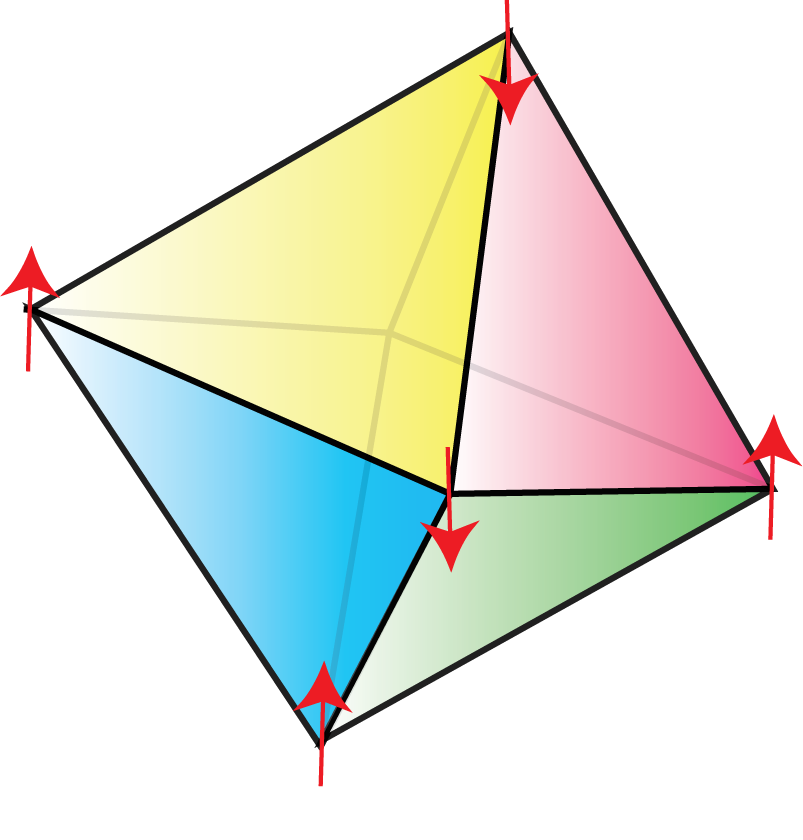}} \quad
\subfigure[Icosahedron]{\includegraphics[width=2.5cm]{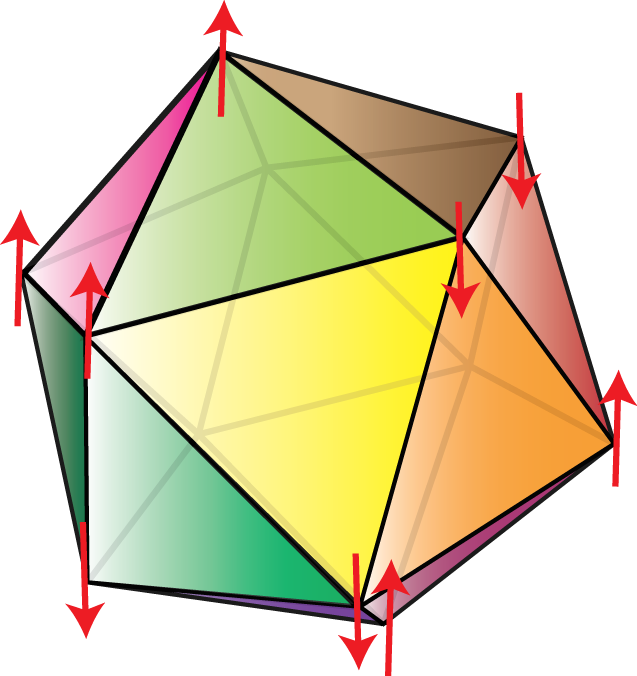}} 
\caption{Ising systems on regular polyhedra}
\label{regular polyhedra}
\end{figure}

The Ising model on the triangular lattice is composed only of triangular plaquettes, and each spin is connected to six neighboring spins. The best small-network approximation to the triangular lattice would be obtained from a polyhedron on which each face is a triangle and each vertex is connected to six neighboring vertices. However, it can be shown~\citep{Lembracht2009} with standard topological methods that no such polyhedron exists. Instead, we consider networks formed from existing regular triangulated polyhedra, namely the tetrahedron, the octahedron and the icosahedron. The geometric structures of these polyhedra are shown in Fig.~\ref{regular polyhedra}.

\begin{figure}[htbp]
\centerline{\includegraphics[width=9cm]{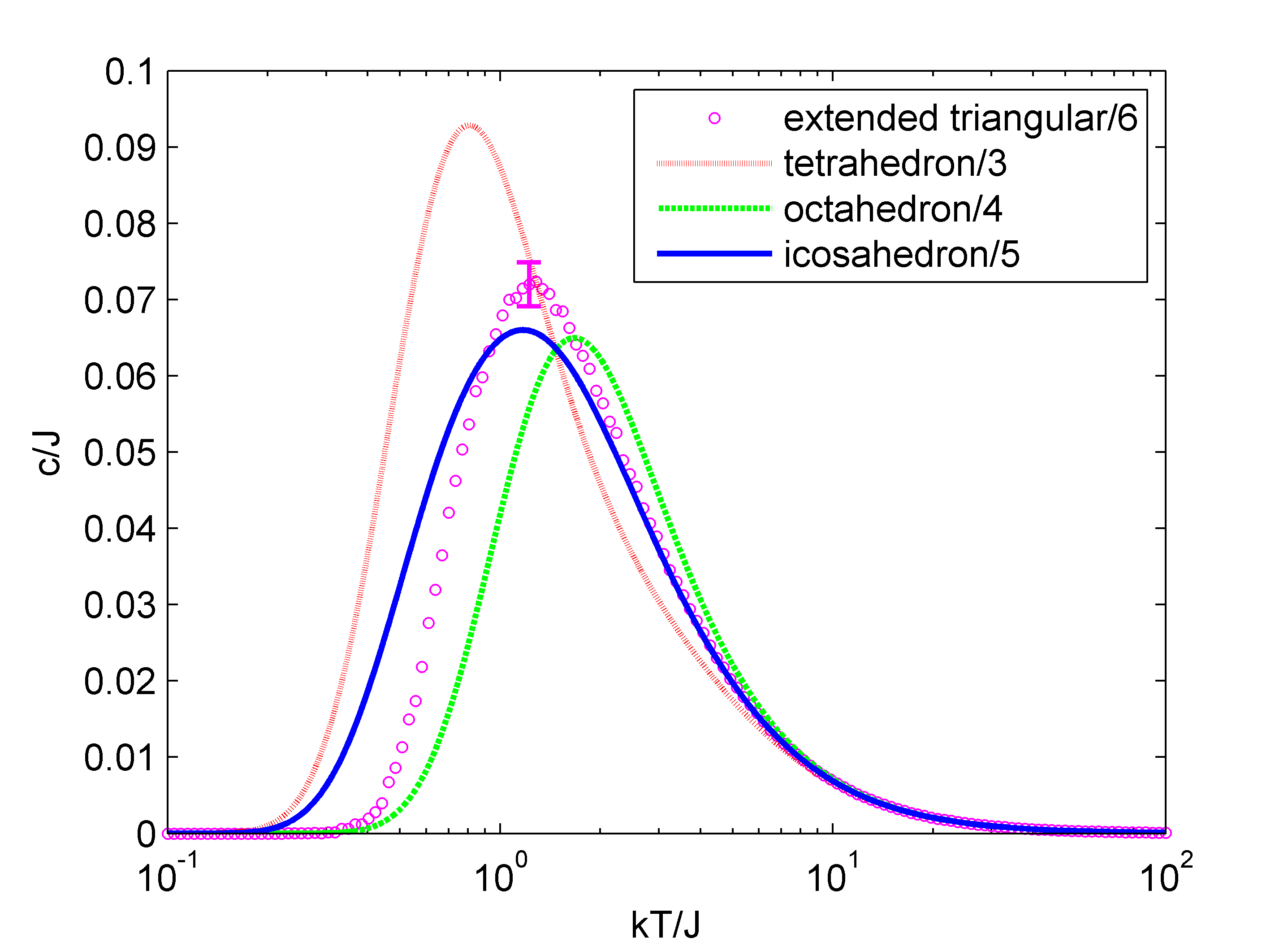}}
\caption{Specific heat of Ising networks on three regular polyhedra and the triangular lattice}
\label{heatC_polyhedra}
\end{figure}

The specific heat profiles of Ising networks on these polyhedra are shown in Fig.~\ref{heatC_polyhedra}. The results show that none of these networks gives a particularly great approximation to the extended triangular lattice. However, as we consider small networks from tetrahedron to octahedron to icosahedron, the approximation gets better. Even though there is no network (that we have considered) that gives an excellent fit, the approximations given by the octahedron and the icosahedron networks are pretty good given that they only have 6 and 12 spins respectively and can therefore be exactly solved with minimal computational power.

\begin{figure}[htbp]
\centerline{\includegraphics[width=3cm]{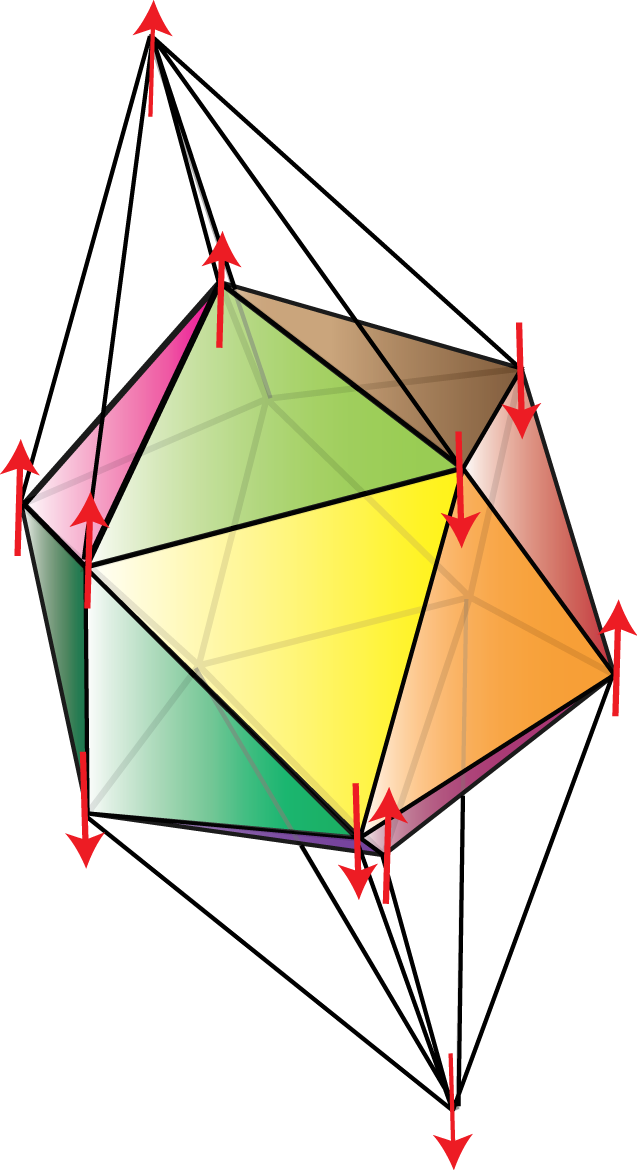}}
\caption{Ising system on the ``icosahedron+2'' network, an icosahedron with two added spins such that every spin has six neighboring spins}
\label{icosahedronPlus2}
\end{figure}

\begin{figure}[htbp]
\centerline{\includegraphics[width=9cm]{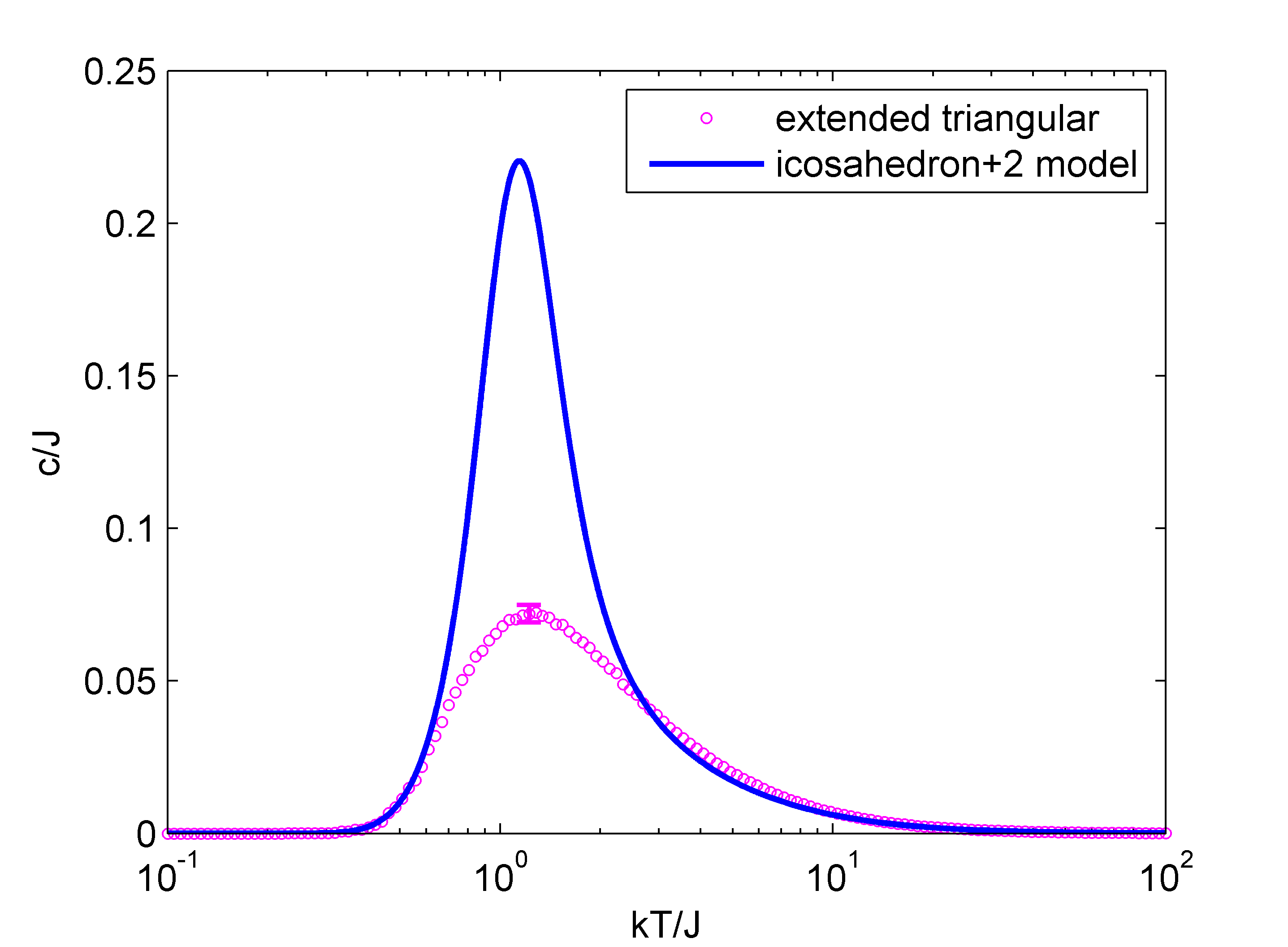}}
\caption{Specific heat of Ising networks on an icosahedron with two added spins and the triangular lattice}
\label{heatC_icosaPlus2}
\end{figure}

The regular polyhedra presented above do not have the same number of neighbors for each spin as the triangular lattice. To find out whether a better approximation can be achieved by a structure in which every spin has six neighbors, we consider an ``icosahedron+2'' network, constructed by adding two spins to the icosahedron network as shown in Fig.~\ref{icosahedronPlus2}. The resulting network has 14 spins in total and every spin in the structure has six neighbors. In Fig.~\ref{heatC_icosaPlus2}, the specific heat profile for the ``icosahedron+2'' network shows that its behavior is very far from that of the triangular lattice, although both structures are made up of triangular plaquettes and have 6 neighbors to each spin. This deviation is due to the arrangement of the triangular plaquettes: while every bond is shared by two triangular plaquettes in the triangular lattice, some bonds in the ``icosahedron+2'' network are shared by three or five triangular plaquettes. This results in a higher density of triangular plaquettes and a corresponding increase in frustration. 

\begin{figure}[htbp]
\centerline{\includegraphics[width=3cm]{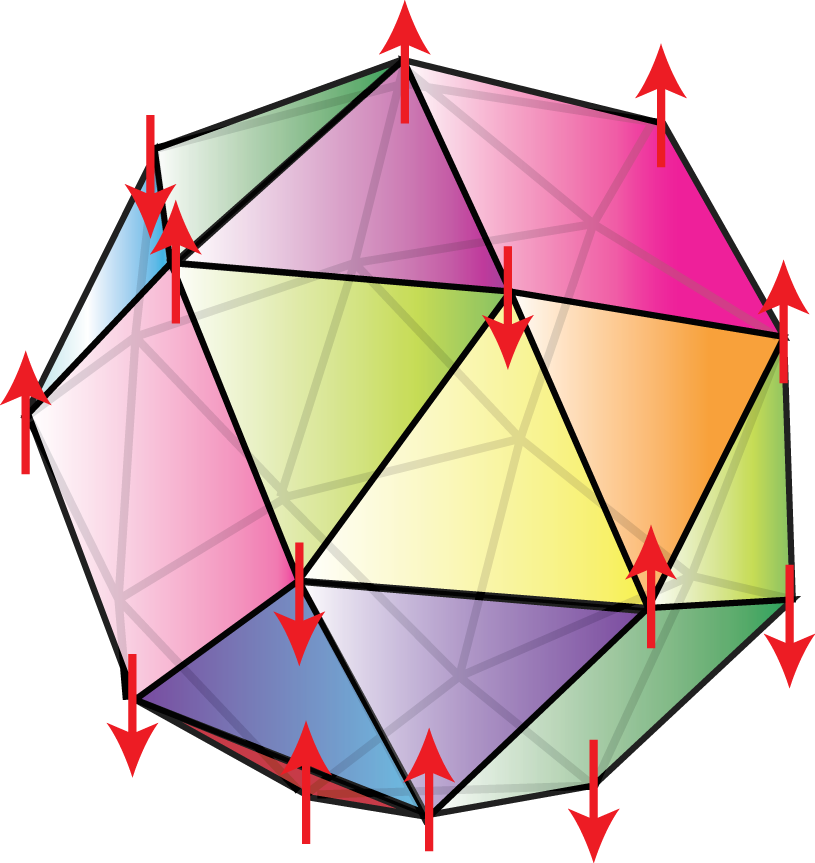}}
\caption{An Ising system on the snub cube}
\label{snubcube}
\end{figure}

\begin{figure}[htbp]
\centerline{\includegraphics[width=9cm]{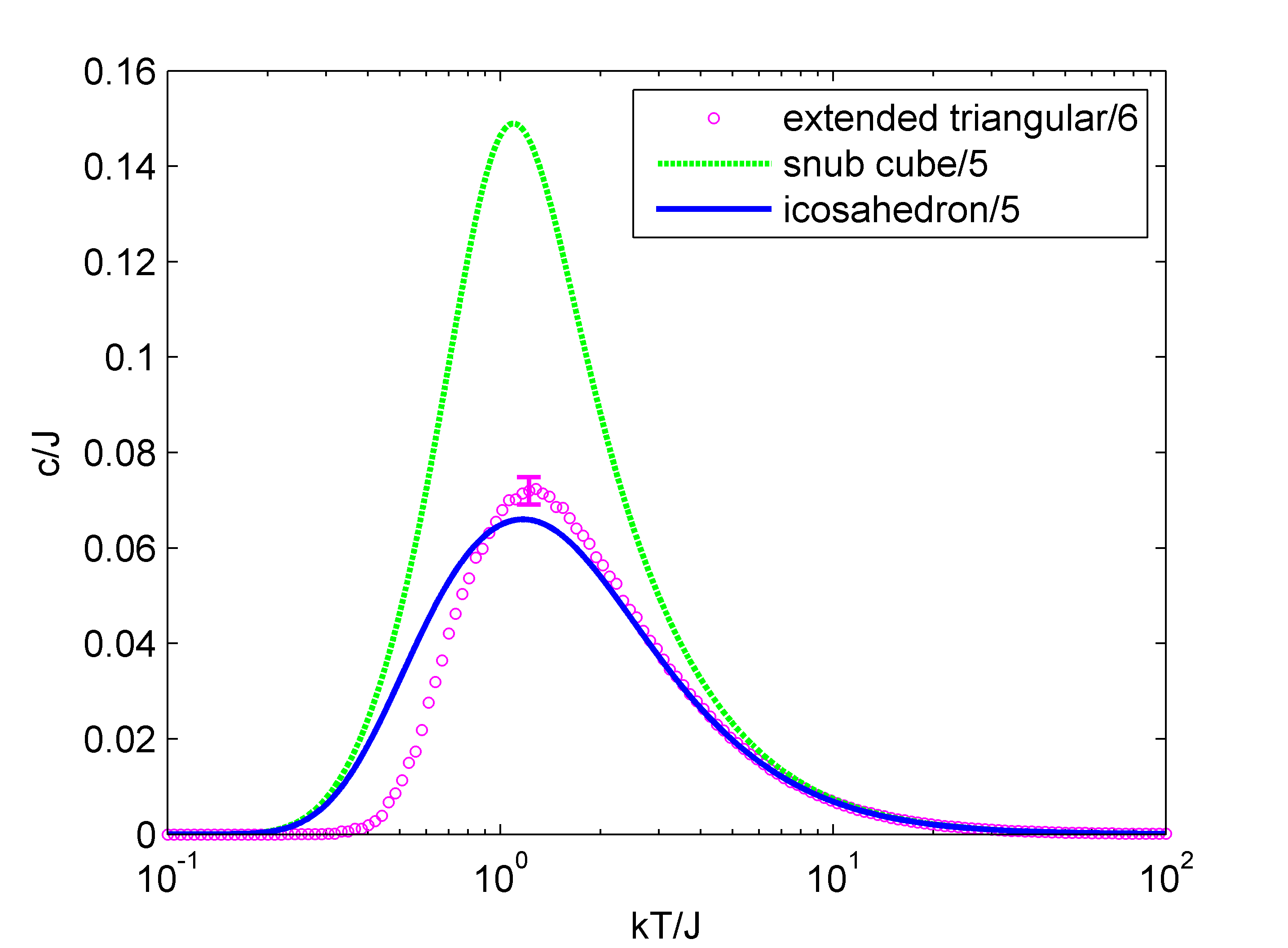}}
\caption{Specific heat of Ising networks on the snub cube, icosahedron and the triangular lattice}
\label{HeatC_icosa_snubcube}
\end{figure}

To check whether a small network has to be made of triangular plaquettes in order to be a good approximation to the triangular lattice, we have also considered a network on the snub cube, whose structure is shown in Fig.~\ref{snubcube}. This structure has 24 vertices and 60 bonds. Each vertex in the structure has five neighboring vertices (as in the icosahedron), but its surface consists of squares and triangles. The specific heat of the frustrated Ising network on the snub cube is shown in Fig.~\ref{HeatC_icosa_snubcube} together with that of the triangular lattice and the icosahedron network. We find that the snub cube network is a much worse approximation to the triangular lattice than the icosahedron network, which also has five neighboring spins to each spin but is made up of triangular faces only. This result suggests that a good approximation-network is obtained for the triangular lattice only if the small structure is made up exclusively of triangular plaquettes.

\subsection{Summary and Comparison of Small-Network Approximations}

In Table \ref{tab:Deviation Index}, we tabulate the deviation index $D$  for the various small Ising networks considered in the previous sections, calculated according to Eq.~\eqref{R_sq}. The value of $D$ provides us with an indicator of the how well each small network approximates its corresponding extended lattice. A small network can be considered a very good approximation if $D>0.9$. For the TKL, we calculate $D$ for systems with $J_{aa}/|J_{ab}|=9$. The values of $D$ suggests that the triangular Kagome lattice is very closely approximated by the triangular drying-rack network, and that the Kagome lattice is very well approximated by the cuboctahedron, the icosidodecahedron and the bowtie-networks. Furthermore, the icosahedron network, which has only 12 spins, serves as a reasonably good approximation to the triangular lattice.  Drawing on our results for the specific cases of the triangular, Kagome, and triangular Kagome lattices, in the next section we summarize general criteria for constructing small networks that are good approximations to extended frustrated systems.

\begin{table}[htbp]
	\centering
		\begin{tabular}{llcc}
		\hline\hline
		extended lattice \qquad & small system \qquad & \quad $N$ \quad & \quad $D$ \quad \\ 
		\hline\hline
		TKL    & triangular drying rack       & $9$  & $1.0000$  \\
		\hline
		Kagome lat.    & cuboctahedron        & $12$ & $0.9909 $ \\
		               & icosidodecahedron    & $30$ & $0.9988 $ \\
		               & 5-bowtie             & $15$ & $0.9986 $ \\
		               & 6-bowtie             & $18$ & $0.9966 $ \\
		\hline
		triangular lat.    & tetrahedron     & $4$ &   $0.4428 $ \\
		                   & octahedron      & $6$ &   $0.8642 $ \\
		                   & icosahedron     & $12$ &  $0.9267 $ \\
		                   & icosahedron$+2$ & $14$ &  $-1.5526 $ \\
		                   & snub cube       & $24$ &  $-0.4429 $ \\
		                   & PBC on $4\times 4$ unit cells & $16$ & $0.3416$\\
		\hline\hline
		\end{tabular}
	\caption{Values of coefficient of Determination $D$ for various small networks}
	\label{tab:Deviation Index}
\end{table}


\section{General Criteria for Constructing Small-Network Approximations}
\label{criteria}

In the previous section, we presented comparisons between a variety of small networks for each of our prototype two-dimensional frustrated Ising systems. A natural question would be: are there any general rules for constructing a small network so that it well-approximates the energetics of a particular extended two-dimensional lattice? Based on our results, we suggest the following general criteria for constructing small networks to model extended two-dimensional lattices:

\vspace{12pt}

\textbf{1. The local lattice structure around each spin in the small network should resemble that in the extended lattice.} 

\begin{figure}[htbp]
\centering
\subfigure[]{\includegraphics[width=1.5cm]{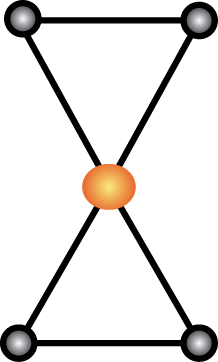}} \qquad
\subfigure[]{\includegraphics[width=2cm]{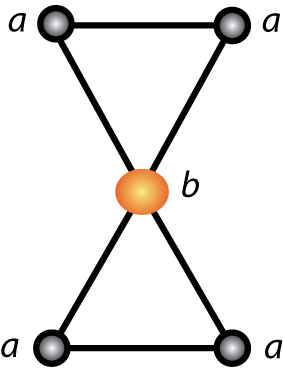}} \qquad
\subfigure[]{\includegraphics[width=2.5cm]{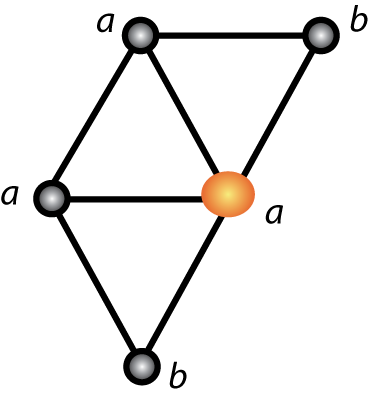}} 
\caption{Local lattice structures around (a) each spin in the cuboctahedron, the icosidodecahedron, the $N$-bowtie networks and the Kagome lattice, (b) each $b$-spin in the triangular drying-rack network and the TKL, and (c) each $a$-spin in the triangular drying-rack network and the TKL. The orange spheres denote the center spins and the grey spheres denote the neighboring spins.} 
\label{local structures good}
\end{figure}

By local lattice structure, we mean the arrangement of triangular plaquettes around each spin in the network. For example, each spin in the Kagome lattice is a shared vertex of two disconnected triangular plaquettes, as shown in Fig.~\ref{local structures good}(a). We observe that the better approximations to the frustrated kagome lattice, such as the icosidodecahedron, the cuboctahedron and the $n$-bowtie networks, have this same local structure around each of their spins. Similarly, the triangular drying-rack network gives an excellent approximation to the frustrated TKL, and the spins on the two structures have identical local structures. In both the ``drying-rack'' network and the TKL, the $b$-spins have the same local structure shown in  Fig.~\ref{local structures good}(b) and the $a$-spins have the same local structure shown in  Fig.~\ref{local structures good}(c). 

\begin{figure}[htbp]
\centering
\subfigure[]{\includegraphics[width=2cm]{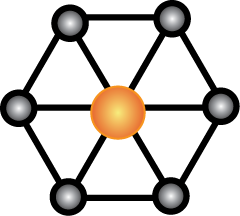}} \qquad
\subfigure[]{\includegraphics[width=2cm]{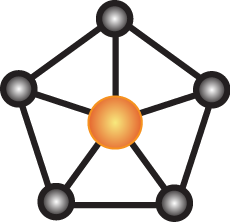}} \qquad
\subfigure[]{\includegraphics[width=2cm]{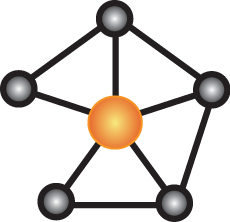}} 
\caption{Local structures around each spin in (a) the triangular lattice, (b) the icosahedron and (c) the snub cube. The orange spheres denotes the center spins and the grey spheres denote the neighboring spins.} 
\label{local structures tri}
\end{figure}

For the triangular lattice, we are not able to find a small network that has exactly the same local structure as the extended lattice, due to the constraints of topology. However, we observe that a better approximation is obtained as the local structure of the small network approaches that of the triangular lattice. The triangular lattice has a local structure of six connected triangular plaquettes around each spin as shown in Fig.~\ref{local structures tri}(a). Among the polyhedra, the local structure of the icosahedron, with five connected triangular plaquettes around each spin as shown in Fig.~\ref{local structures tri}(b), resembles the triangular lattice to the greatest extent. As a result, the network on the icosahedron gives the best approximation to the frustrated triangular lattice. On the other hand, the snub cube network has a local structure shown in Fig.~\ref{local structures tri}(c), in which not all the triangular plaquettes are connected. As the local structure for the snub cube differs sharply from that of the triangular lattice, the behavior of the snub cube network is very different from the triangular lattice.

We note that the importance of the local arrangement of triangular plaquettes results from the fact that the triangular plaquettes are the main contributors to the frustrated behavior in Ising systems on the triangular lattice, the Kagome lattice, and the triangular Kagome lattice. The triangular plaquettes give rise to a higher density of frustrated bonds than other $(2n+1)$-gons. If a geometrically frustrated lattice is made up of pentagons and other higher-sided polygons, then the local arrangement of the pentagons would be the most important factor to consider.

\vspace{12pt}
\textbf{2. Each bond in the small network should be a shared edge of the same number of triangular plaquettes as that in the extended lattice.}

In the lattices that we have considered, every edge is shared by no more than two triangles due to the constraints of two-dimensionality. When constructing small networks for these two-dimensional lattices, we find it is extremely important to make sure that every bond is shared by the same number of triangles as in the corresponding extended lattice. For example, in the Kagome lattice, every bond belongs to only one triangular plaquette. In the small networks which give good approximations for the Kagome lattice (including the icosidodecahedron, the cuboctahedron and the bowtie networks), every bond also belongs to only one triangular plaquette. The same holds true for the triangular drying-rack network model for the TKL.

On the other hand, if this condition is not satisfied, the behavior of the small network deviates sharply from that of the the extended system. For example, even though the number of nearest neighbors are the same in both the triangular lattice and the ``icosahedron+2'' network, the later gives a $D$ value as poor as $-1.5$.  The chief reason is that there are bonds that are shared by three or five triangles, which leads to extra frustration due to the excess connectivity among those triangular plaquettes. On the other hand, in the snub cube network, there are bonds that belong to one triangle only, and this also causes the behavior of this small network to deviate sharply from that of the triangular lattice. 

While the first criterion on the similarity of local structures should be satisfied as closely as possible, the second criterion that the bonds must be shared by the same number of triangles must be satisfied for all small networks. Based on our observations, the behavior of a frustrated lattice is heavily dependent on the types of bonds in the network. Whether a bond is shared by one, two, or more triangular plaquettes makes a drastic difference, as the connectivity of triangular plaquettes through the bonds has a significant effect on the density of states and the number of accessible states in the system. For example, in the ground states of the snub cube, the square plaquettes are not frustrated, and thus the snub cube is less frustrated than the icosahedron, even though both structures have the same number of neighbors for each spin. The snub cube thereby has a lower energy per spin in the ground state, resulting in a much higher peak in the specific heat than the icosahedron (whose behavior is close to that of the frustrated triangular lattice).

\section{Small Networks vs. Periodic Boundary Conditions}
\label{PBC}

In Section \ref{criteria}, we presented general guidelines for constructing small networks to approximate the thermodynamics of extended two-dimensional lattices. Our degree of success is directly related to the short correlation lengths in these systems, which allow local structure to dominate the thermodynamics. Given this, the reader may be wondering whether just as good of an approximation can be obtained by taking a small piece of the two-dimensional lattice and applying periodic boundary conditions (PBC). If the correlation length is zero outside of a unit cell (as it is for the triangular Kagome lattice), applying periodic boundary conditions on one unit cell would be expected to give a very good approximation. In fact, the network obtained from a unit cell of the triangular Kagome lattice with PBC is identical to what we have called the triangular drying-rack network. However, applying PBC to the Kagome and triangular lattices, which have longer correlation lengths, is met with much lower success, as we show next.

\begin{figure}[htbp]
\centerline{\includegraphics[width=9cm]{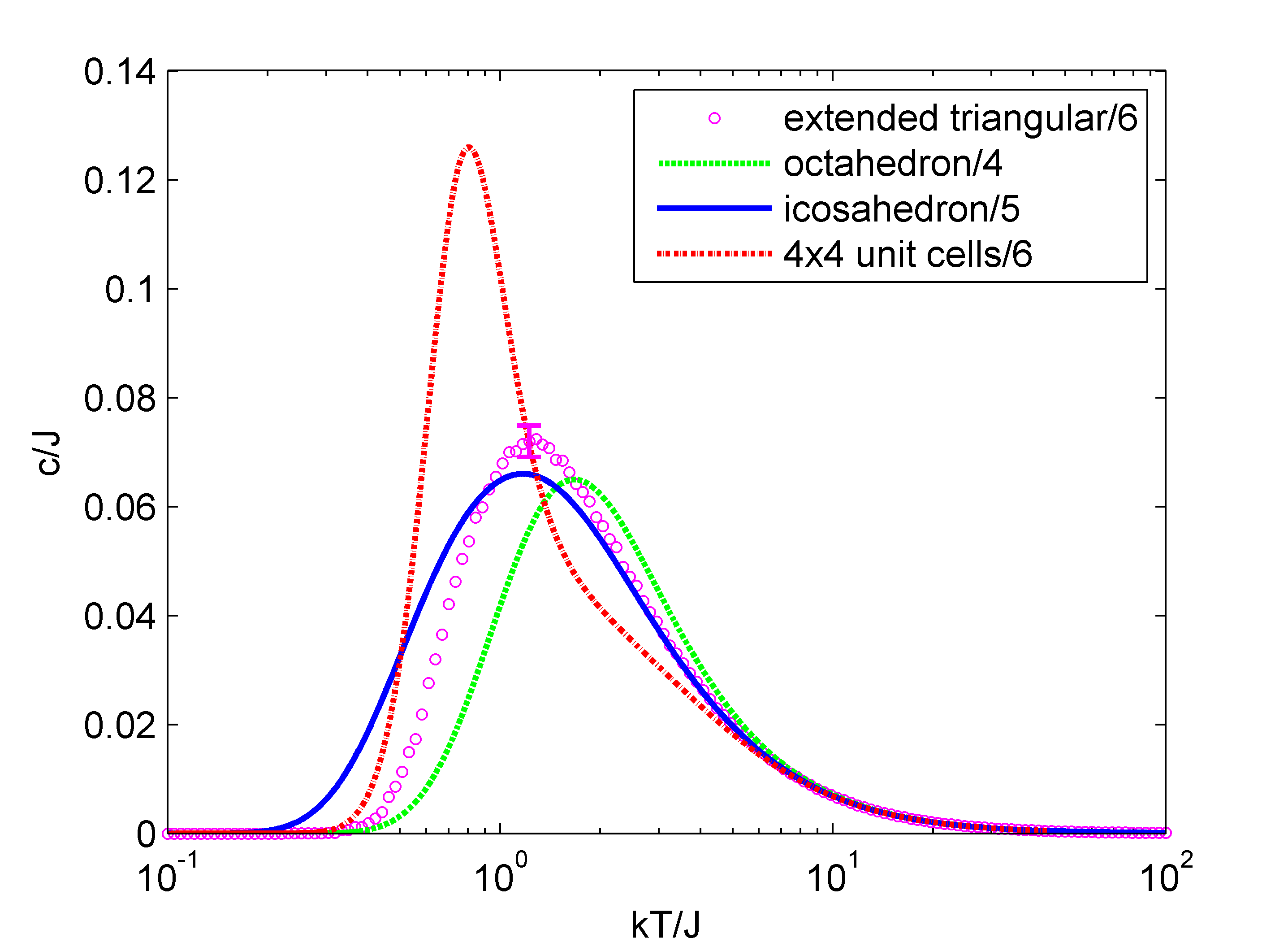}}
\caption{The specific heat profile obtained by applying periodic boundary condition on $4 \times 4$ unit cells of the triangular lattice compared with that for the extended triangular lattice, the octahedron and the icosahedron}
\label{TriPBC}
\end{figure}

For extended two-dimensional systems whose spin-spin correlation decays with distance $r$, it is difficult to know \emph{a priori} the minumum size required for a patch of the lattice with PBC to well-approximate an extended system. We now show that (for a comparable number of spins), a network constructed by applying PBC to a small section of the extended lattice gives a much worse approximation than the small networks we have developed. For the triangular lattice, applying PBC on $4\times 4$ unit cells (16 spins) is insufficient to produce a good approximation to the extended triangular lattice. Fig.~\ref{TriPBC} shows the specific heat profile obtained from simulations on $4\times 4$ unit cells of the triangular lattice with PBC, as well as the specific heat profiles for the octahedron network, the icosahedron network, and the triangular lattice for comparison. Even though $4\times 4$ unit cells has more spins than the octahedron and the icosahedron, its fit of the triangular lattice has a $D$ value of 0.3416 (Table \ref{tab:Deviation Index}), much worse than the icosahedron and the octahedron. The reason behind this huge deviation is that by applying PBC, one introduces additional triangular plaquettes.

We conclude that, compared to applying PBC to a finite piece of the lattice, the method of small-network approximation that we have developed in Section \ref{criteria} is generally more accurate, for the same number of spins. In some cases, one of our small networks has the same structure as a few unit cells of the extended lattice under PBC. In addition to the case of the TKL and the triangular-drying-rack-network, for the Kagome lattice, the network of the cuboctahedron turns out to be the same as applying PBC on $2 \times 2$ unit cells of the Kagome lattice. However, the variety of structures covered by our small-network approximation is much larger than what can be obtained by applying PBC and in the case of the triangular lattice, the small-network approach allows us to do much better than applying PBC.

\section{Conclusion}
\label{conclusion}

In this work, we have examined the correspondence between geometrically frustrated Ising systems on particular small spin networks and three ordinary extended two-dimensional systems. The small correlation lengths in these frustrated Ising systems makes it possible for a remarkably small network of spins to give a good approximation for the specific heat of the corresponding extended two-dimensional lattice. However, the correlation length poses a fundamental lower limit on the size of the small network required to obtain a good approximation, and we find decreasing success for this small-network approach as the correlation length of the extended lattice system increases. We suggest that using suitably-designed small networks is a good way to obtain a first approximation for the properties of large and complicated geometrically frustrated systems. 

The behavior of frustrated magnetic systems are often difficult to study, particularly beyond the simplest approximations. For many physical systems, Heisenberg spins are required to accurately model the quantum properties of the systems. Even within the classical Ising approximation, computationally-intensive Monte Carlo simulations on large systems are often used because there are no general analytic approximations available for frustrated systems. In this work, we have presented an approach that may provide a general way to approximate frustrated systems with extremely small computational expense. The small networks discussed here could be embellished to resemble more realistic models, for instance by using Heisenberg spins instead of Ising ones, while remaining computationally feasible. In addition, the idea of small-network approximations could lead to new renormalization or numerical procedures. In this work, we have focused on ordinary two-dimensional systems that are frustrated due to their triangular plaquettes. However, similar methods could be applied to other kinds of geometrically frustrated lattices with various geometrical structures in both two- and three-dimensions, such as spin ice.

\begin{acknowledgments}
This work was supported by the Research Corporation through a Cottrell College Science Award (CL) and by a Jerome A.\,Schiff Fellowship from Wellesley College (BZ).
\end{acknowledgments}


\bibliographystyle{apsrev}
\bibliography{thesis_bibb}

\end{document}